# Measuring the Impact of Interference Channels on Multicore Avionics

Steven H. VanderLeest
Rapita Systems, Inc
Novi, MI, USA
svanderleest@rapitasystems.com

Samuel R. Thompson
Rapita Systems Ltd.
York, UK
sthompson@rapitasystems.com

*Abstract—* **Measurement-based analysis of software timing behavior provides important insight and evidence for flight certification of modern avionics systems. For multicore systems, however, this analysis is challenging due to interference effects from shared hardware resource usage. We present an approach to multicore timing analysis that uses interference generators to stress interference channels in multicore systems. The test methodology comprises two steps. First, platform characterization measures the sensitivity of the hardware and RTOS to targeted interference channels using a combination of interference generators. Second, software characterization measures the timing behavior of instrumented applications while interference is generated on shared resources.**

I. INTRODUCTION

Multicore processing has great potential benefits to reduce Size, Weight, and Power (SWaP) at lower costs. However, that benefit can easily be squandered if interference between cores is not sufficiently mitigated. The focus of this paper is on techniques for accurately measuring the impact of interference on the Worst Case Execution Time (WCET) for flight software.

In this introduction, we cover the basic concepts of multicore processing, including isolation techniques and the importance of measuring the impact of multicore interference channels on WCET. The rest of the paper then covers our V&V framework in section II, our test methodology in section III, and a discussion on the analysis of completeness in section IV. The paper wraps up with a brief description of future directions for our research.

*A. Multicore Processing*

To provide some context, this section reviews the technological shift to multiple cores on a chip, techniques to isolate independent functionality, and relevant flight certification standards.

*1) Trend from Unicore to Multicore*

In 1965, Gordon Moore stated his eponymous law that the number of transistors on an integrated circuit doubles approximately every 18 months. Moore's Law has shown remarkable longevity, bearing reasonably true to this doubling period over the five decades since it was introduced. For the first several decades, the CPU clock frequency also tended to have exponential growth, reaching 1 GHz at the turn of the century, and true to form, 2 GHz about 1.5 years later. However, since that time, limitations in power dissipation have limited maximum frequency. While transistor count has continued to grow exponentially, today, very few chips have clock speeds exceeding 4 or 5 GHz. To continue meeting customer performance needs, chip producers have begun using multicore solutions. A multicore processor contains two or more processing units on a single chip. For example, the Xilinx Zynq UltraScale+ MPSoC (Multi Processor System on Chip) contains four Cortex-A53 processors in the Application Processor Unit (APU) and two Cortex-R5 processors in the Real-time Processor Unit (RPU), all on the same piece of silicon. The processors in a multicore system typically share much of the rest of the computing hardware, including most of the memory and I/O subsystems. This complicates control and adds a significant challenge to assurance analysis because software applications can now run simultaneously on different cores and thus might interfere with one another more directly than on unicore systems where partitions must run one at a time. Mechanisms to isolate applications and reduce interference are examined in the next subsection.

*2) Isolation Techniques*

Isolation is an architectural mechanism that separates software functions so that they do not interact or affect one another in any significant way. Also known as separation or partitioning, isolation prevents one software function from impacting the behavior of other independent software functions present on the system, so that each software function deterministically meets its requirements regardless of the presence or absence of others. Isolation must deterministically limit interference between partitions so that even in the worst case, no partition uses more than its allocated share of each resource within a given time period. Isolation is a safety and security property of the system that contributes to certification evidence by demonstrating that each partition is guaranteed to receive the resources it needs to complete its tasks within the required time.

Partitioning has two main aspects: temporal and spatial. Temporal partitioning mechanisms control the sharing of resources (e.g., the CPU) by time-sharing access by allocating the resource to one software partition at a time and then switching access to other partitions according to some deterministic schedule. Spatial partitioning mechanisms control the sharing of resources (e.g., main memory) by dividing them into more granular sub-components and allocating each sub-component to a particular partition with exclusive access. Thus, temporal partitioning allocates the entire resource for a fraction of time while spatial partitioning allocates a fraction of the resource for the entire time. Bandwidth partitioning mechanisms combine these two types, sharing a resource by allocating a certain amount of space per time, such as maximum message bytes per time frame. Most partitioned avionics systems use temporal partitioning to share the CPU, spatial partitioning to share the memory, and bandwidth partitioning to share the I/O. However, this is not the only option. For a multicore processor, for example, cores may be partitioned spatially, but each core may then also be partitioned temporally.

Isolation is essential to obtain deterministic, predictable behavior from partitioned functions. Isolation that can be verified as rigorous separation of partitions in an Integrated Modular Avionics (IMA) system provides significant advantages including modularity, ease of integration, and reduced certification effort for mixed-criticality systems. Partitioning provides the benefit of modularity and portability. Software functionality mapped to partitions and with interfaces restricted to the ARINC 653 API are generally compile-time portable to other ARINC 653 operating systems. The divide-and-conquer philosophy of modularity also simplifies the system architecture design by breaking up a large, complex system into smaller, and thus more understandable, units. Partitioning provides the benefit of ease of integration by eliminating unanticipated interactions between added components. Partitioning reduces certification costs, especially for mixed-criticality systems. Without partitioning, all software in the system, even low criticality applications, must be certified to the level of rigor of the most critical application, significantly raising Verification and Validation (V&V) costs. Isolation reduces future certification efforts because the introduction of a new partition does not require a complete repetition of V&V testing on the existing partitions. That is, the design assurance of the isolation itself justifies the reduced need for testing for interaction between the partitions. However, this implies that the assurance for the isolation must be at a high level of rigor.

*3) Certification Guidelines and Standards*

The Federal Aviation Administration (FAA) in the US oversees civilian flight certification (and military airworthiness authorities frequently adopt similar standards). The requirements for assuring avionics designs are extensive and arduous, especially for high Design Assurance Levels such as A and B. The FAA indicates in Advisory Circular 20-115C that "DO-178C is an acceptable means of compliance for the software aspects of type certification." In turn, DO-248C provides supporting information to DO-178C. In describing partitioning, DO-248C mentions ARINC 653 as an example. Furthermore, DO-297 notes ARINC 653 in a design example as a means by which a system designer might "ensure portability and robust partitioning of applications." Partitioned operating environments following the ARINC 653 standard are the most common type of RTOS for IMA.

The MULCORS [1] research study conducted for the European Aviation Safety Agency identifies flight certification issues related to multicore COTS platforms. The US Federal Aviation Administration (FAA) released the CAST-32 position paper on flight certification of systems with multicore processors in 2014 and then produced an updated CAST-32A [2] document in 2016 with an intent "to identify topics that could impact the safety, performance, and integrity of a software airborne system executing on Multi-Core Processors." According to the FAA website, a new Advisory Circular (AC) harmonized with an Acceptable Means of Compliance (AMC) from the European Union Aviation Safety Agency (EASA) is anticipated, stating that "publication of A(M)C 20-193 is planned using CAST-32A content"[3]. The current position paper notes that multicore processors have the benefit of greater computational power than uniprocessors, allowing multiple software applications to run simultaneously. However, this benefit can also come at the cost of unforeseen interference for shared resources. Thus, the effect of interference on determinism and safety is the primary focus of the paper. The paper defines an "interference channel" as any "platform property that may cause interference between independent applications".

*B. Importance of WCET for Flight Certification*

Whether submitted to military or civilian airworthiness authorities, a key piece of evidence supporting flight certification is the accurate estimation of Worst-Case Execution Time (WCET), which is defined as the maximum time that a task requires on the flight computing hardware. Software domains that are not safety-critical do not usually consider WCET, instead aiming to optimize the average software execution time. In contrast, for real-time safety-critical domains, we must ensure that the application will complete its critical tasks within the allotted time budget under all possible scenarios, even the worst cases. While WCET estimation can be challenging even for the application running alone on the hardware, the analysis task can become significantly more complex when other software applications are simultaneously running on the system.

*1) How WCET can be absurdly large with multicore*

A naïve first impression of performance on multicore might be to anticipate that the performance on a four-core system will be 4x the performance on a single core. However, the reality is that applications running on each core can interfere with each other. In the extreme case, if all the cores are attempting to use the same limited-bandwidth shared resource simultaneously, the resulting performance can be abysmally low, potentially even worse than the single core performance. For example, in one study, a finely tuned denial-of-service attach on a shared structure in cache slowed the baseline application by 300x [4]. Practically, the ideal of 4x performance can only be approached when the applications running on distinct cores are using distinct resources or using shared resources in a strictly partitioned manner.

WCET due to multicore interference is complex, depending on the resource usage and timing patterns of each application, the nuances of the operating environment, the configuration of the hardware, and the fine (and sometimes undisclosed) details of the processor architecture. This complexity makes it almost impossible to model, and thus predict, the impact that using multiple cores will have on the WCET of each application. Instead, a measurement-based approach must be used to achieve reasonable insight into the WCET impact.

*2) Interference Generator approach to measuring WCET*

The measurement approach we present below is based on the use of an interference generator, which we call a RapiDaemon, to stress targeted shared resources, measuring the resulting impact on WCET. We have suggested the use of interference generators in the past [5], as have others. Guet suggested generating parametric interference to the data cache [6], and Hwang notes the use of a virtual machine as an interference generator [7]. Iorga refers to their interference generators as "enemy programs" [8]. A group at Barcelona Supercomputing Center uses specialized microbenchmarks to put a high load on a targeted shared resource to measure interference [9]. One of the new contributions of this paper is to place these methods within an overall framework.

## II. VERIFICATION AND VALIDATION FRAMEWORK

### A. Context: Importance of Well-Structured Verification

It is of critical importance to provide independent verification for any claim made regarding multicore timing. This extends not only to claims of WCET estimation but also to any wider claims, such as claims of efficacy of RTOS-provided partitioning mechanisms. All mechanisms to mitigate interference – whether in hardware, the RTOS, or anywhere else – must be independently verified.

The verification framework must be well structured but also sufficiently flexible so that it can effectively be used for the full gamut of multicore certification activities on both current and future platforms. All the activities undertaken should be portable to bare-metal systems to "bare RTOS" configurations and to full 653-based IMA platforms; indeed the framework is currently being actively applied to systems in all these categories for aerospace clients. Through the MASTECS project [10] we are working with key industry partners to address this full spectrum of platform types.

To provide an appropriate and versatile structure to the design, implementation, and reporting of multicore timing analysis activities, a multicore timing analysis V-framework has been designed as shown in Figure 1.

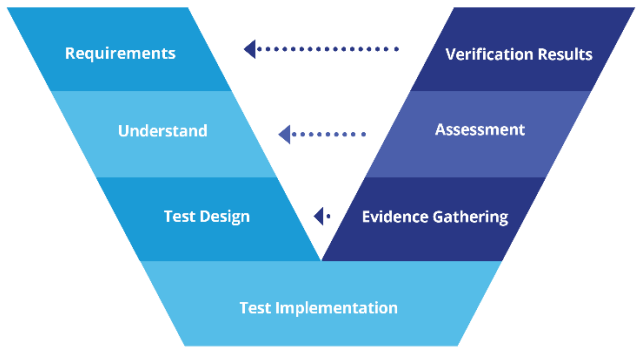

*Figure 1: V-model testing framework.*

The framework is iterative – there are many points at which it is appropriate to take the outputs from one stage of the process and feed them back into an earlier stage of the process. Examples of these feedback steps are included in the following explanation of the model.

### B. Multicore Timing Analysis V-Framework

The following section comprises a description of each of the steps of the V-model framework, including elucidation of the specific CAST-32A objectives satisfied by the different stages.

*1) Requirements*

The first step in the multicore verification and validation process is the requirements specification. This is effectively a scoping activity; requirements may have a mapping to those from the DO-178C process ("This task shall complete within a specified deadline") or may be independent. During a platform selection activity, for example, requirements could be based upon the effectiveness of partitioning provided by the platform. For the purposes of multicore timing analysis, the platform consists of the combination of both hardware and software.

Multicore timing analysis can represent a significant body of work; it is important to ensure that the requirements specified at this point in the process maintain a narrow scope. When writing requirements it is also of value to consider what the criteria are for satisfaction of a requirement. An example of a poor requirement with ill-defined stopping criteria could be "The multicore platform shall have bounded interference." The stopping criterion for this requirement would be having exhaustively tested every possible access pattern to every possible combination of all available resources; this is computationally unfeasible. Worse, as processors can contain undocumented resources [11]; to strictly satisfy this requirement, there would be an implicit requirement to demonstrate that all possible resources have been identified. This is also computationally unfeasible. A more achievable requirement might be "The software under test shall have bounded impact on WCET in the presence of multicore interference measured for the set of channels identified by analysis to be significant".

*2) Understand*

After requirements have been enumerated, it is necessary to develop a good understanding of the relevant hardware and software and build up some hypotheses regarding system operation. This allows the scope of the analysis to be further restricted. For example, if it were determined by inspection that contention on a specific resource could not affect the software (for example, software which doesn't require use of an particular peripheral should be unaffected by contention on it), then no further measurements concerning usage of the resource would be necessary. This is in agreement with the guidance for CAST-32A objective MCP_Resource_Usage_3, which states: "If the applicant identifies interference channels that cannot affect the software applications in the intended final configuration, then those interference channels do not need to be mitigated and no verification of mitigation is needed."

In another example, if it had been demonstrated using interference generators that a partitioning system prevents any measurable interference on a shared L2 cache, then it is reasonable to argue that it is unnecessary to carry out any specific tests on the software under test concerning L2 cache interference. Again, this is congruent with CAST-32A objective MCP_Software_1, which states that for platforms with robust partitioning, applications may be verified independently.

Similarly, if it can be demonstrated that the software under test will never use a resource, then it may be maintained that the impact of interference on that resource need not be experimentally confirmed for that software. However, care should be taken to ensure that no invalid assumptions are made. For example, if testing a piece of software in isolation shows that the software fits entirely onto a shared L2 cache and never makes requests for data from main memory, it would be invalid to assume that interference on main memory is therefore irrelevant for this software. This is because is it possible that other tasks executing on other cores could evict data from the shared cache and lead to the task relying upon data fetched from main memory in the final integrated system. More detailed discussion on the matter of identifying shared resources which may be of importance is given in Section II.C.

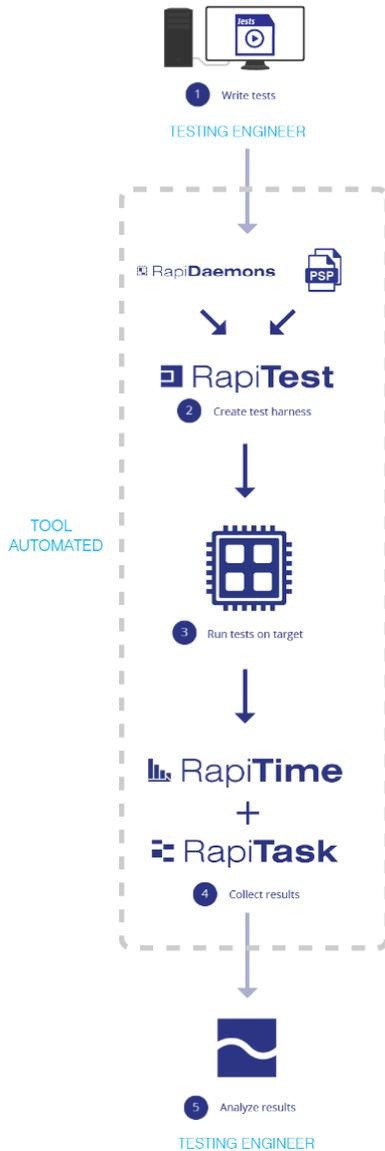

*Figure 2: Test implementation workflow as supported by the Rapita Verification Suite (RVS)*

*3) Test Design*

When the scope of the investigation has been restricted as far as possible, a test or suite of tests sufficient to verify the requirement(s) should be designed. Of particular importance is the definition of 'success criteria' against which the results will be assessed; these criteria should be selected to ensure that they are necessary and sufficient to satisfy the corresponding requirement. For a detailed discussion of the philosophy behind the design of multicore tests, see Section III.

Designing tests which measure applications while they are suffering interference from RapiDaemons on different resources provides direct evidence in support of CAST-32A objective MCP_Software_1.

*4) Test Implementation*

Implementation of the tests is a naturally necessary step. This includes the integration of the testing framework and tooling with the build system and embedded target, and then use of the testing framework in implement the tests themselves.

When implementing tests, it is worth considering the traceability of test implementations within the context of the wider testing regime by ensuring that traceability into any requirements management system is maintained. A possible workflow for this stage of the framework is illustrated in the diagram in Figure 2.

*5) Evidence Gathering*

The execution of tests is broadly similar to that of any other on-target testing. Notwithstanding, there are a few crucial differences that may manifested depending on the tests that are being executed.

*a) Tracing multiple cores simultaneously*

For some multicore timing tests, it is necessary to extract execution traces from multiple cores simultaneously. For example, to measure the timing behavior of some software under test on one core while simultaneously verifying that the correct RapiDaemon interference generators are running at the correct times on other cores. This implicitly requires that timing data can be synchronized between cores as well.

*b) Reading performance monitoring counters*

Virtually all processors provide some built-in performance-monitoring tools that can be used to provide highly granular visibility of resource usage (for example, dirty evictions from L2 cache). These can be useful for a range of activities, among which many are multicore-specific or multicore-relevant. For example, before a RapiDaemon interference generator can be used for carrying out timing tests, its behavior must be verified, and performance counters can be very useful for this verification.

A good example of the usefulness of performance counters is preventative scheduling. If two tasks are known to access the same resource intensively enough that they cause each other to miss their respective deadlines, one so-called safety-net solution could be to modify the schedule to prevent the two tasks from accessing the resource at the same time. To determine the effectiveness of this scheduling policy, it may be necessary to record resource usage on more than one core at once and ensure that the readings are time-synchronized.

*c) Feedback*

Test execution and data gathering is also an early feedback point for the design. Exposition on the matter is included below:

It may be that basic runtime behavior necessitates immediate modification of experimental configuration or revision of underlying assumptions. One real-world example from a customer project was using an 8-core platform with an RTOS; when an L2 cache RapiDaemon interference generator was instantiated on a particular core, the entire RTOS stalled until the RapiDaemon was killed.

There could also be problems with the traces extracted. Many issues with the test implementation can be identified in this way. Examples could include tests whose execution traces are not parseable because test implementations made assumptions about the behavior of threading, or tests that rely upon the availability of kernel-space functionality from user-space code and return null values rather than useful data.

Finally, some RTOSes/hypervisors contain built-in functionality to limit the potential for misbehaving software to impact the rest of the system. This could include the automated throttling of resource bandwidth to 'greedy' applications, which could impact the efficacy of RapiDaemons. Similarly, applications that have missed RTOS-enforced deadlines because of RapiDaemon interference could be killed by the RTOS, thus preventing the trace from containing direct evidence of the magnitude of any interference. This is the stage at which such problems can be identified, allowing the configuration of the platform to be modified.

*6) Assessment*

Results analysis is the process of turning captured timing traces into a form that can be used to assess the correctness of any hypotheses of system operation and to assess the satisfaction or otherwise of tests using the previously-defined success criteria. For the most part, this step can be automated using a testing tool, e.g., the Rapita Verification Suite (RVS). If the success criteria and test implementations have been well-designed, this should be a simple case of comparing the results with the success criteria.The third and final element of results analysis is feedback stage. oThe principal reasons for feedback at this stage are poorly-conceived success criteria that cannot be properly assessed using the results available from the tests. This could necessitate either clarification of the success criteria to ensure that the results captured from the target are sufficient to fully satisfy them, or modification of the tests themselves to generate results that can be used to assess the success criteria.

*7) Verification Results*

The reporting stage is where reports that are necessary to support certification are generated. Depending on the workflow, this could be a semi-automated process whereby the toolchain generates template artifacts that have been pre-populated with the results from the previous stage.

It is critically important to ensure that the artifacts produced are sufficient to satisfy all the requirements imposed by the relevant regulatory authority regarding the units that have been tested. Depending on the chosen structure for certification artifacts, the evidence could either constitute part of the Software Accomplishment Summary (SAS) or could be treated separately in a multicore appendix. This satisfies CAST-32A objective MCP_Accomplishment_Summary.

This is the final step at which feedback can occur within the overall framework. Feedback at this stage is generally triggered by the inadequacy of the final results to fully address the requirements. Feedback at this stage is more expensive, as it will generally require that the entire process be iterated.

*C. Identifying Resources Shared by Cores*

The exhaustive identification of resources shared between cores is not a trivial task. This task starts with the requirements stage and is pursued in earnest in the understand stage of our framework. A starting point could be block diagrams of the hardware, as most shared resources can be identified from this high-level hardware overview. However, this view lacks nuance. For example, many modern multicore processors provide each core with private L1 caches for data and instructions. A naïve approach might suggest that, since these caches are private to individual cores, they cannot constitute an interference channel. This is not true; consider the case that the L1 caches of two cores reference the same location in memory and one of the cores makes a modification that invalidates the cached copy of that data in the other private L1; it is necessary for the cache coherency mechanism to access ('snoop') the other L1 to ensure that the stale cache line is invalidated. Even though the caches are nominally private, it is nevertheless possible for them to be affected by cross-core interference.

Other non-obvious modes of interference may also exist, particularly when applications are using more than one resource simultaneously, and this can create couplings between different interference channels. Equally, the underlying architecture of the hardware may have a non-obvious impact. For example, a shared cache may be composed of several underlying banks of memory – accesses from different cores that access the same physical bank of memory may suffer increased latency, while simultaneous accesses to cache lines on different banks may not.

### III. TEST METHODOLOGY

When designing tests using our V-framework, a four-step methodology can be used to provide the necessary output to satisfy CAST-32A objectives relating to the characterization of interference channels and the verification of software timing behavior. This methodology tests a single interference channel at a time; it should be repeated for each interference channel under consideration.

To implement this methodology successfully, it is essential to have access to some means of measuring the resource usage to a useful degree of granularity. For this purpose, access to relevant performance monitoring counters on the target hardware is necessary. Note that static analysis methods cannot be used alone for this purpose; particularly when there is an RTOS in use, statically predicting the behavior of mechanisms such as network-on-chip interconnects and pseudo-random caching policies is not a solvable problem.

At the same time, performance counters are generally intended to be little more than debugging aids and are rarely properly tested by silicon vendors; it is common to find that performance counters will misbehave by counting the wrong thing, counting incorrectly, or not counting at all.

Before trusting performance counters for a certification project, they must, therefore, be tested to verify that they are counting correctly. Special RapiDaemons can be used to perform this counter verification activity.

*A. Platform Characterization*

The first half of the testing methodology concerns the characterization of the target platform (consisting of the hardware and RTOS (if used) in their final intended configuration). This comprises two steps detailed below:

*1) Victim Running in Isolation*

The execution time behavior of a RapiDaemon is measured while the RapiDaemon is running on one core of the target system and nothing is executing on any of the other cores. This is used as a baseline against which other measurements can be compared.

*2) Victim Running Against Adversaries*

Additional RapiDaemons are instantiated on other cores on the target system. These additional RapiDaemons should access the same resource as the instrumented RapiDaemon; this configuration attempts to generate contention on the target resource(s).

If this interference channel is present in the platform, the expected outcome is that the execution time behavior of the instrumented RapiDaemon will be affected in some way. Generally speaking, there are two forms that this modification of behavior can take:

1. The victim takes longer to execute.
2. The victim execution time becomes less predictable.

The two effects above may be seen in combination and with various magnitudes. The example in Figure 3 shows a victim main memory RapiDaemon on an NXP T2080, running alone (in green) and running against three adversary main memory RapiDaemons (in red).

If no effect is seen from executing with the additional RapiDaemons, this suggests that the interference channel is not present in the platform – if a RapiDaemon that utilizes a resource very heavily cannot be interfered with by other RapiDaemons accessing the same resource very intensively, it is possible to argue that a real application generating far less intensive resource accesses will not interfere with other real applications.

This means that in addition to addressing CAST-32A objective MCP_Resource_Usage_3 by determining which resources should be addressed by further testing, this step can be used to narrow the scope of the investigation and remove the requirement for testing many interference channels. This decreases the necessary testing effort and test duration, lowering the cost of certification.

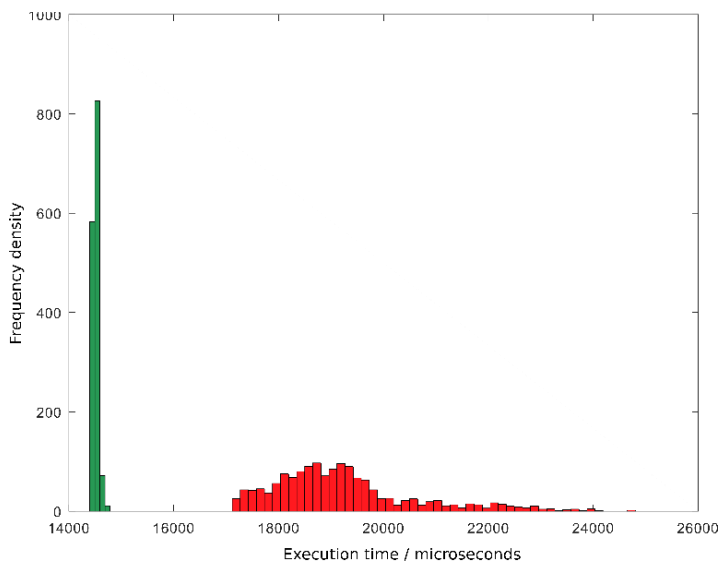

*Figure 3: Histogram execution time profile*

Another outcome of this stage is the measurement of the efficacy of any resource partitioning mechanisms in place on the target, provided by either the hardware or any RTOS. If RapiDaemons accessing a resource that is supposedly partitioned are unable to impact their execution time behavior, then it may be concluded that the resource partitioning is indeed robust as described in CAST-32A.

*B. Software Characterization*

When the characteristics of the underlying platform are well-understood, the software to be hosted on the platform can then be investigated. It is crucial that when carrying out testing on the software that it is adequately exercised to give confidence that the execution time behaviour has been determined for all operational circumstances.

This is a two-step process as follows:

*1) Fingerprint Application*

The purpose of this step is to allow the investigation of the behavior of the software under test to be restricted to the minimal useful scope. For example, if it were to be determined that a piece of software never made any accesses to a particular resource when executing normally, then it would not be necessary to test against RapiDaemons generating load on that resource.

Once the resource usage profile for a code item has been gathered, this profile can be used to determine on which resources it should be tested for response to contention. For example, if the peak use rate of a given resource were found to be less than that seen in a RapiDaemon configuration that didn't yield any interference, then it is reasonable to conclude that no effect is likely to be seen for this software; therefore further testing against that resource could be deemed unnecessary.The information gained in this step can also be used for the optimization of the application to make the best use of the known available resources in the system.

The best insights into the application behavior are obtained by instrumenting the application source code. However, in cases where the source code is not available, the behavior can still be detected via black box testing and use of external performance measures, e.g., observing processor Performance Monitor Counters or using a bus analyzer.

*2) Application Running Against Adversaries*

Having determined which resources are most important for assessing the execution time of the software running on the target platform, the software should then be executed against RapiDaemons that generate load on these resources.

The target application is executed on one core, while RapiDaemons are executed on the other available core(s). Exemplar results from this activite are shown in Figure 3. The execution time measured when the RapiDaemons are also running is an upper bound on WCET for the software under test because the RapiDaemons cause equivalent or higher interference than any real application could cause.

For this step, the application can be treated as a black box, i.e., it is not necessary to have the source code order to measure the interference impact. However, if the impact is not acceptable, instrumentation of the application can be helpful in optimizing the system for better isolation of independent applications.

## C. Analysis of Results

The test steps described above provide a distribution of observed WCET, which can be used to estimate the actual WCET of the system. In cases where the increase in WCET due to multicore interference is negligible, the system requirements are likely still met with sufficient margin. In cases where the WCET increases dramatically and requirements are no longer met, iteration on our test framework can provide insights in order to optimize timing, such as identifying sections of code experiencing the highest impact or varying the hardware and RTOS critical configuration settings to improve isolation.

## IV. COMPLETENESS OF INTERFERENCE CHANNEL CHARACTERIZATION

CAST-32A notes that the complexity of multicore systems can lead to many different forms of interference, all of which must be identified and mitigated to demonstrate airworthiness. Thus, our analysis of interference must not only be rigorous for each channel, but we must also rigorously demonstrate the completeness of our list of channels. The task is similar in nature to the work determining whether the set of system requirements is complete – it requires human wisdom and peer review to provide sufficient assurance of completeness.

A previous paper suggests that a taxonomy of system architectural features and isolation mechanisms can provide a conceptual checklist that contributes to our confidence of completeness [12]. However, it must be adapted to specific cases, as every new system design may potentially contribute an entirely new architectural feature or novel isolation technique. Even existing hardware and software used in new ways may produce a new interference channel. Thus, thorough analysis with independent peer review is necessary as evidence of completeness. This may not be sufficient, however, requiring timing measurements on target to confirm the underlying assumptions of the analysis. Complete hardware information regarding the mechanisms by which cores access shared resources is also necessary. This may not be sufficient, however, since documentation sometimes contains errors, again pointing to the need for confirmation by measurement and test.

## V. FUTURE WORK

We continue active research on this topic to refine our processes and tools to provide a package for compliance with CAST-32A. Our future work will include consideration of model-based approaches while still incorporating measurement for validation and consideration of best practices for determining the cumulative WCET.

## A. Measurement and Modeling

Model-Based Development (MBD) is often used to quickly produce code from high-level models. MBD tools that have been designed with formal methods and/or fully qualified may at some point produce code that is correct by design and requires little or no further verification. However, even in these cases, the isolation mechanisms that the system provides to segregate applications are an assumption of the MBD tools that must itself be validated in support of the assurance evidence of the application. As discussed earlier, the difficulty in fully modeling complex multicore hardware and interactions of the RTOS pushes us toward the continued need for on-target measurement of WCET for multicore interference channels. Nevertheless, there may be some benefit to enhancing or calibrating MBD tools with the results of such measurements.

## B. Cumulative WCET

Another area of future work is to standardize the means of accumulating results across combinations of interference on interference channels. Although current guidance indicates that all relevant interference channels must be analyzed, it is not clear how combinations of interference channels might produce unique impact that cannot be anticipated by the individual contributions. For example, if the WCET increase due to multicore interference is 5% for shared resource A and 7% for shared resource B, but the resources A and B are accessed by the cores through entirely independent interconnects, then the appropriate margin must be the maximum of the two, 7%. However, if the two resources are accessed over a common bus, the appropriate margin may be the sum of the two, 12%, or even higher if contention between the two types of resources produces timeouts or other collision responses. We hope to better describe a test methodology, measurement process, and analysis technique to produce a reliable estimate of the total margin of safety needed for WCET in the presence of simultaneous interference channels.

## VI. CONCLUSION

It is still early in the history of certification of fully active multicore systems hosting mixed-criticality software, but thus far techniques for verification such as we propose seem to be the most promising way to provide solid certification evidence to assure that multicore interference has been sufficiently mitigated.

ACKNOWLEDGMENT

This project has received funding from the European Union's Horizon 2020 research and innovation program under grant agreement No 878752.